\documentclass{article} \usepackage[round]{natbib}
\usepackage{epsfig,fullpage,graphicx,url} 
\def\x{{\mathbf x}}
\newcommand{\citenoun}[1]{{\citet{#1}}} 
\newcommand{\cut}[1]{}

\title{Detecting influenza outbreaks by analyzing Twitter messages}

\author{Aron Culotta\\
Department of Computer Science\\
Southeastern Louisiana University\\
Hammond, LA 70402}

\begin{document}

\maketitle
\begin{abstract}
We analyze over 500 million Twitter messages from an eight month
period and find that tracking a small number of flu-related keywords
allows us to forecast future influenza rates with high accuracy,
obtaining a 95\% correlation with national health statistics. We then
analyze the robustness of this approach to spurious keyword matches,
and we propose a document classification component to filter these
misleading messages. We find that this document classifier can reduce
error rates by over half in simulated false alarm experiments, though
more research is needed to develop methods that are robust in cases of
extremely high noise.
\end{abstract}

\section{Introduction}
\label{sec:intro}
There has been growing interest in monitoring disease outbreaks using
the Internet. Previous approaches have applied data mining techniques
to news articles
\citep{grishman02information,mawudeku06global,brownstein08surveillance,reilly08indications,collier08bio,linge09internet},
blogs \citep{corley10text}, search engine logs
\citep{eysenbach06info,polgreen08using,ginsberg09detecting}, and Web browsing
patterns \citep{johnson04analysis}. The recent emergence of {\sl
  micro-blogging} services such as Twitter.com presents a promising
new data source for Internet-based surveillance because of message
volume, frequency, and public availability.

Initial work in this direction includes \citenoun{ritterman09using},
who show that Twitter messages can improve the accuracy of market
forecasting models by providing early warnings of external events like
the H1N1 outbreak. More recently, \citenoun{quincey09early} have
demonstrated the potential of Twitter in outbreak detection by
collecting and characterizing over 135,000 messages pertaining to the
H1N1 virus over a one week period, though no attempt is made to
estimate influenza rates.

Two similar papers were recently published that estimate national
influenza rates from Twitter messages
\citep{lampos10tracking,culotta10towards}. Both use linear regression
to detect keywords that correlate with influenza rates, then combine
these keywords to estimate national influenza
rates. \citenoun{lampos10tracking} train and evaluate on a much larger
data set (28 million messages) than used in
\citenoun{culotta10towards} (500K messages), which likely contributes
to the differing quality of the estimates (97\% correlation with national
statistics on held-out data in \citenoun{lampos10tracking}, 78\%
correlation in \citenoun{culotta10towards}).

In this paper, we report results of our analysis of over 570 million
Twitter messages collected in the 8 months from September 2009 to May
2010. This data was originally collected as part of the work of
\citenoun{oconner10from}, in which a strong correlation is revealed
between certain Twitter messages and political opinion
polls\footnote{We are extremely grateful to Brendan O'Connor for
  sharing this data with us.}.

The contributions of this paper are as follows:
\begin{itemize}
\item We find that simple keyword matching produces a surprisingly
  high correlation with national statistics. For example, the
  proportion of Twitter messages containing the word ``flu'' produces
  a 84\% held-out correlation with weekly influenza-like-illness
  statistics reported by the U.S. Centers for Disease Control and
  Prevention. Adding of a handful of other flu-related
  terms improves this to 95\%.
\item Despite these strong correlations, we find that the methodology
  of selecting keywords based on their correlation with national
  statistics is flawed because of the high likelihood of detecting
  false correlations. For example, the phrase ``flu shot'' has a
  correlation greater than 90\%, but certainly this is not a good term
  to monitor, as it may spike in frequency without a corresponding
  spike in influenza rates. Similar problems arise from events such as
  drug recalls and flu-related policy announcements, which receive a
  lot of discussion in social networks, but are not necessarily
  indicative of an increase in flu rates.
\item We propose a method to estimate robustness to false alarms by
  simulating false outbreaks like those described above. Using this
  measure, we show that by adding a document classification component
  to remove spurious keyword matches, we can reduce the severity of
  false alarms while preserving accurate forecasting estimates.
\end{itemize}

In Section \ref{sec:data}, we first describe the national influenza
statistics as well as the Twitter dataset. Then in Section
\ref{sec:corr}, we describe the methodology of correlating Twitter
messages with national statistics, and report correlations on a range
of keyword combinations. In Section \ref{sec:spurious}, we discuss the
impact of spurious keywords on correlation results. In Section
\ref{sec:classifier} we introduce and evaluate a document classifier
to filter spurious messages. Finally, in Section \ref{sec:simulation}
we provide results of false alarm simulations, finding that document
filtering reduces mean-squared error by over half, while maintaining a
94\% correlation with national statistics on held-out data.

\section{Data}
\label{sec:data}
We begin with a description of the data used in all experiments.

\subsection{Influenza Monitoring in the United States}
\begin{figure}[t]
\centering
\begin{tabular}{cc}
\epsfig{file=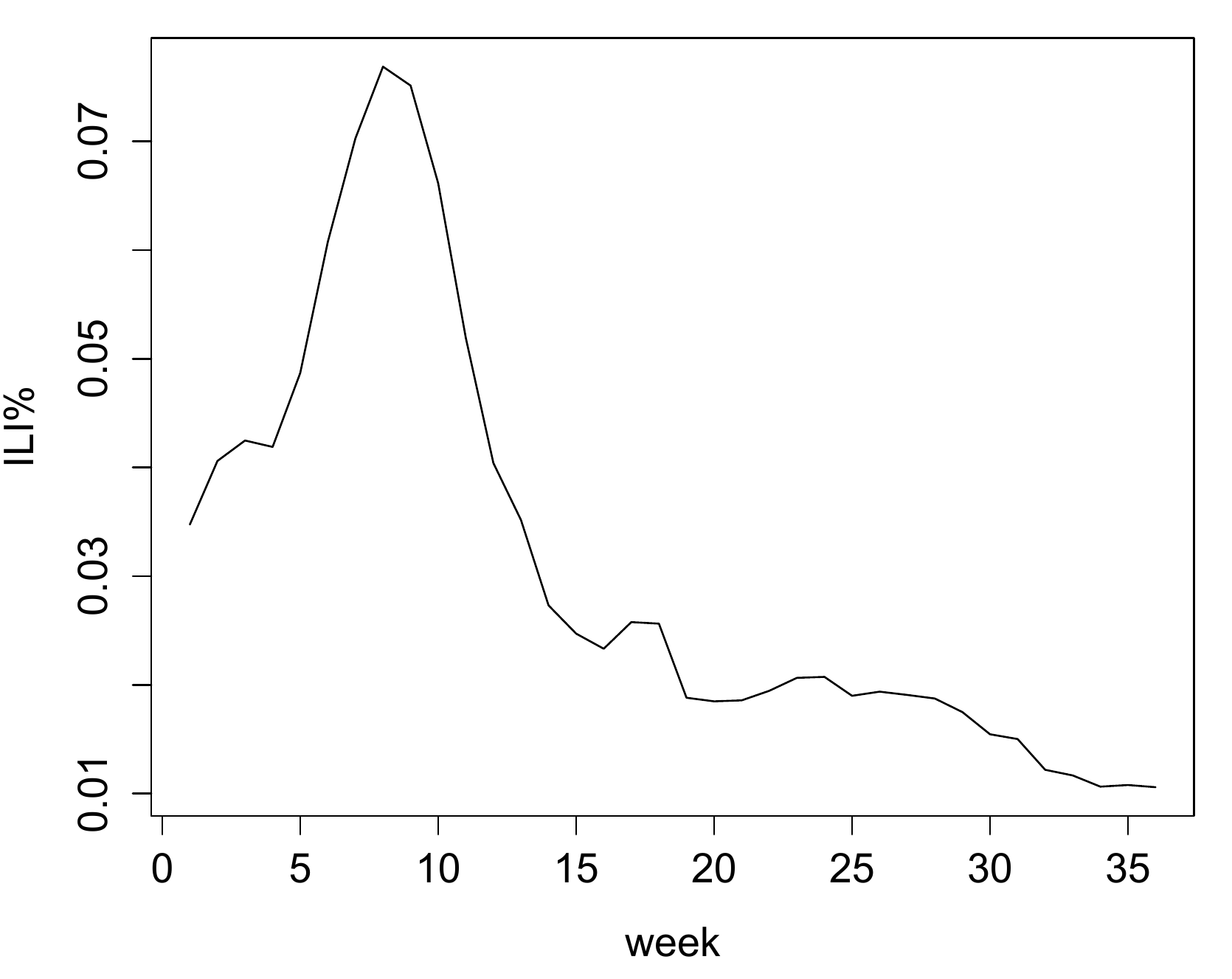,width=0.5\linewidth,clip=} & 
\epsfig{file=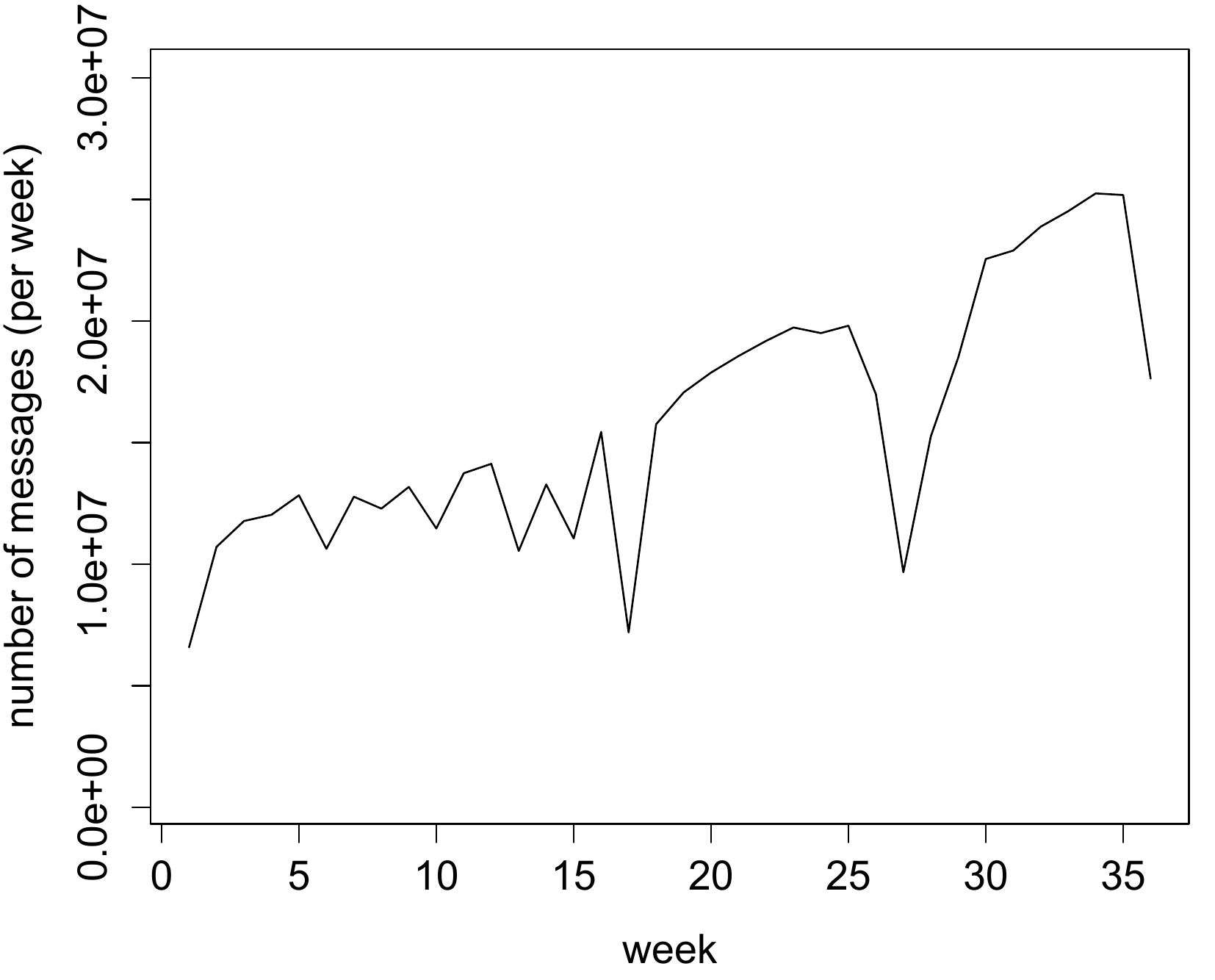,width=0.5\linewidth,clip=} \\
\end{tabular}
\caption{The left figure shows the ILI rates as obtained form the CDC's weekly tracking statistics. Week 1 ends on September 5, 2009; week 36 ends on May 8, 2010. The right figure displays the number of Twitter messages collected per week over the same time period. \label{fig:ili}}
\end{figure}

The U.S. Centers for Disease Control and Prevention (CDC) publishes
weekly reports from the US Outpatient Influenza-like Illness
Surveillance Network (ILINet). ILINet monitors over 3,000 health
providers nationwide to report the proportion of patients seen that
exhibit influenza-like illnesses (ILI), defined as ``fever
(temperature of 100$\,^{\circ}$ F [37.8$\,^{\circ}$ C] or greater) and
a cough and/or a sore throat in the absence of a known cause other
than
influenza.''\footnote{\url{http://www.cdc.gov/flu/weekly/fluactivity.htm}}
Figure \ref{fig:ili} shows the ILI rates for the 36 week period from
August 29, 2009 to May 8, 2010.

While ILINet is a valuable tool in detecting influenza outbreaks, it
suffers from a high cost and slow reporting time (typically a one to
two week delay). The goal of this line of research is to develop
methods that can reliably track ILI rates in real-time using Web mining.

\subsection{Twitter Data}
Twitter.com is a micro-blogging service that allows users to post
messages of 140 characters or less. Users can subscribe to the {\sl
  feeds} of others to receive new messages as they are written. As of
April 2010, Twitter reports having 105 million users posting nearly 65
million message per day, with roughly 300,000 new users added
daily \citep{oreskovic10twitter}.

There are several reasons to consider Twitter messages as a valuable
resource for tracking influenza:
\begin{itemize}
\item The high message posting frequency enables up-to-the-minute analysis of an outbreak.
\item As opposed to search engine query logs, Twitter messages are
  longer, more descriptive, and (in many cases) publicly available.
\item Twitter profiles often contain semi-structured meta-data (city,
  state, gender, age), enabling a detailed demographic
  analysis.
\item Despite the fact that Twitter appears targeted to a young
  demographic, it in fact has quite a diverse set of users. The
  majority of Twitter's nearly 10 million unique visitors in February
  2009 were 35 years or older, and a nearly equal percentage of users
  are between ages 55 and 64 as are between 18 and
  24.\footnote{{\sl Twitter older than it looks}.
    Reuters MediaFile blog, March 30th, 2009.}
\end{itemize}

The Twitter messages used in this paper are a subset of those used in
\citenoun{oconner10from}, restricted to the 2009-2010 flu season from
September 2009 to May 2010. \citenoun{oconner10from} gathered the
messages through a combination of queries to Twitter's public search
API as well as messages obtained from their ``Gardenhose'' stream, a
sample of all public Twitter messages.

Figure \ref{fig:ili} shows the number of Twitter messages obtained per
week for the same time frame as the ILI percentages. The average
number of messages per week is 15.8 million. Due to difficulties in
data collection, there are a few anomalies in the data, but even the
smallest sample (week 1) contains 6.5 million messages.

\begin{figure}[t]
\centering \includegraphics[height=5.2in,width=6in]{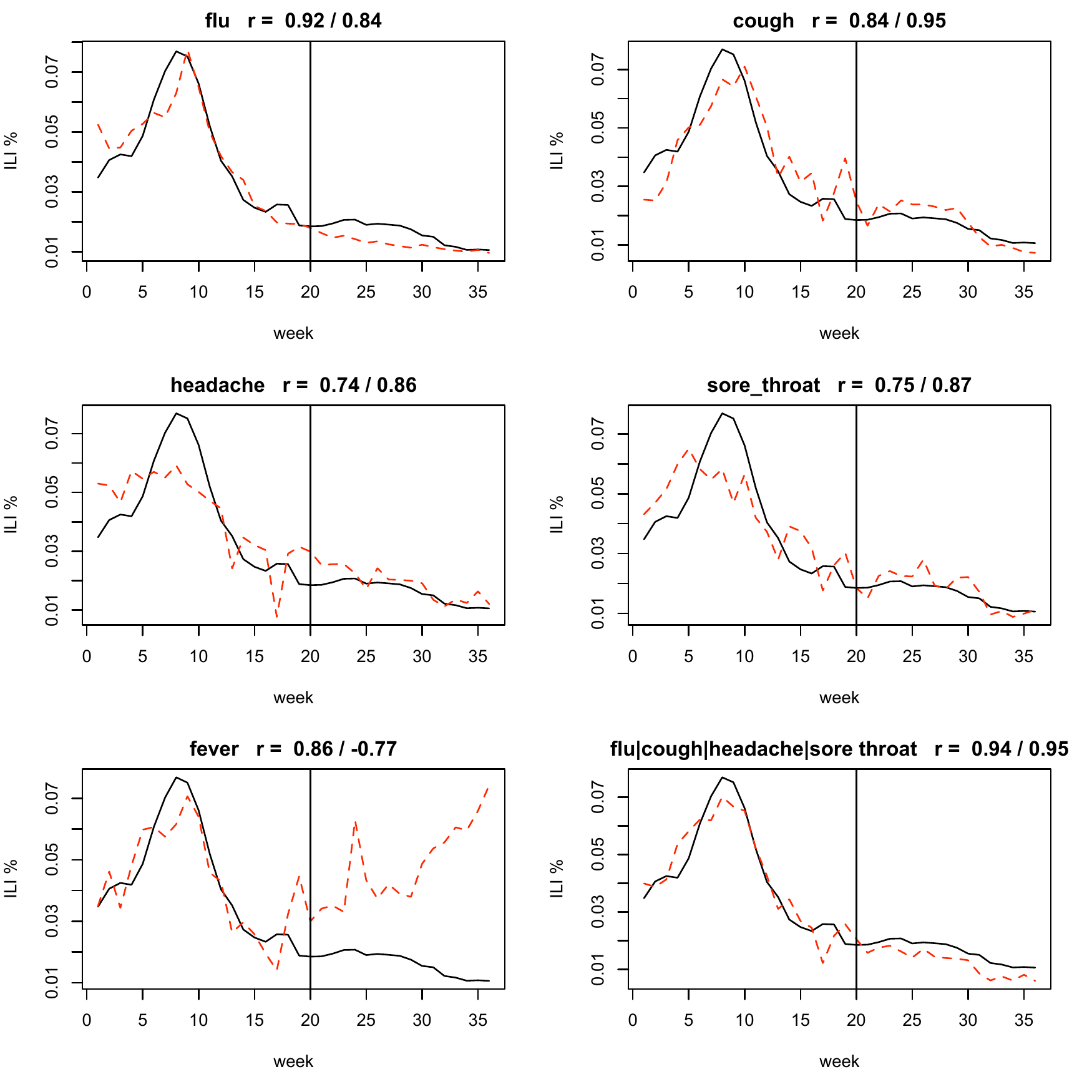}
\caption{Fitted and predicted ILI rates using regression over query fractions of Twitter messages. Black lines are the true ILI rates, red lines are the fitted/predicted values, with the vertical line splitting the training and evaluation data.  \label{fig:kw}}
\end{figure}

\section{Correlating keywords with ILI rates}
\label{sec:corr}
In this section, we describe a simple methodology to correlate Twitter
messages with ILI rates. Let $P$ be the true proportion of the
population exhibiting ILI symptoms. In all experiments, we assume $P$
is the value reported by the CDC's ILINet program.

Let $W = \{w_1 \ldots w_k\}$ be a set of $k$ keywords, let $D$ be a
document collection, and let $D_W$ be the set of documents in $D$ that
contain at least one keyword in $W$. We define $Q(W,D) =
\frac{|D_W|}{|D|}$ to be the fraction of documents in $D$ that match
$W$, which will refer to as the {\sl query fraction}.

Following \citenoun{ginsberg09detecting}, we first consider
a simple linear model between the log-odds of $P$ and $Q(W,D)$: 
\begin{equation}
\label{eq:regression}
\hbox{{\sl logit}}(P) = \beta_1 \hbox{{\sl logit}}(Q(W,D)) + \beta_2 + \epsilon
\end{equation} 
with coefficients $\beta_1, \beta_2$, error term $\epsilon$, and logit
function {\sl logit(X)} $=\hbox{ln}(\frac{X}{1-X})$.

Figure \ref{fig:kw} displays the result of this regression for a
number of keywords. We fit the regression on weeks 1-20, and evaluate
on weeks 21-36. In each figure, the black line is the true ILI rate,
the red line is the fitted/predicted rate. The vertical line indicates
the transition from training to evaluation data. The title of each
plot indicates the query used as well as the training and evaluation
correlation values.

These results show extremely strong correlations for all queries
except for {\sl fever}, which appears frequently in figurative phrases
such as ``I've got Bieber fever'' (in reference to pop star Justin
Bieber).  

Because these results are competitive with those in
\citenoun{lampos10tracking}, who use a more complex methodology, the
conclusion we draw from these results is that even extremely simple
methods can result in quite accurate models of ILI rates from Twitter
data.

\section{Analysis of spurious matches}
\label{sec:spurious}
While these strong correlations are encouraging, we must be very
careful about the conclusions we draw. Many of these correlations may
be spurious. For example, a number of messages containing the term
``flu'' are actually discussing ``flu shots'', ``flu vaccines'', or
are simply referencing news stories about the flu. While these type of
messages may correlate with ILI rates, they are likely not the types
of messages researchers have in mind when they report these
correlations. Instead, the system would ideally track mentions of
people reporting having the flu or flu-like symptoms, as opposed to
simply mentioning the flu in passing.

{\bf These spurious correlations leave keyword-based methods extremely
  vulnerable to false alarms.} For example, a recall of a flu vaccine,
a governmental policy announcement regarding flu, or a release of a
new flu shot will all lead to a spike in messages containing the word
flu.

\begin{figure}[t]
\centering 
\includegraphics[height=5.2in]{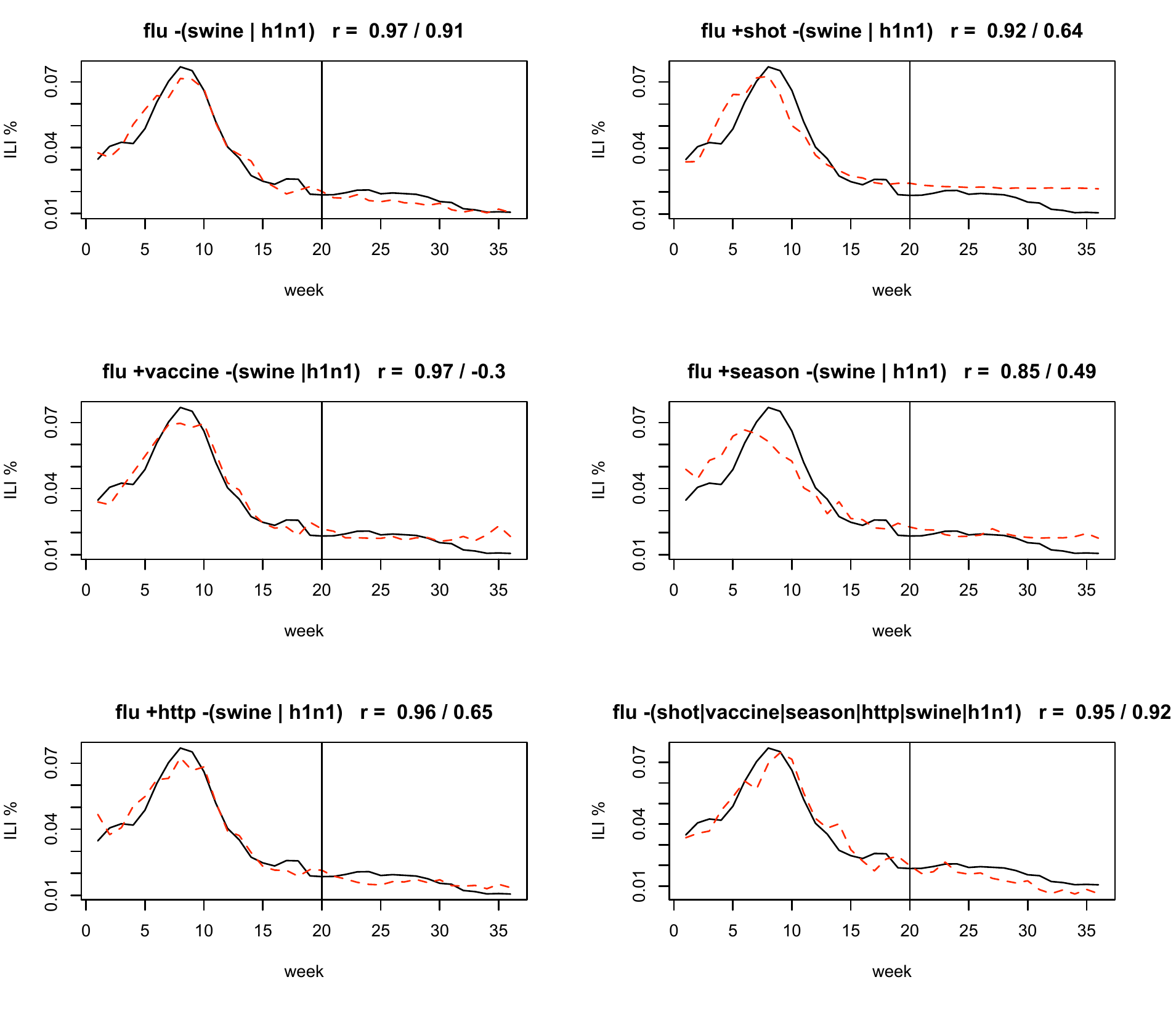}
\caption{Correlation results with refinements of the flu query. \label{fig:kw2}}
\end{figure}

\begin{table}[t]
\begin{center}
\begin{tabular}{|c|c|}
\hline
{\bf query} & {\bf \# messages returned}\\
\hline
\hline
{\bf flu} & 387,405 \\
\hline
{\bf flu +(swine  h1n1)} & 198,149\\
\hline
{\bf flu +shot} & 37,010 \\
\hline
{\bf flu +vaccine} & 35,102\\
\hline
{\bf flu +season} & 11,850\\
\hline
{\bf flu +http} & 166,123\\
\hline
\bf flu -(swine h1n1 shot vaccine season http) & 126,194 \\
\hline
\end{tabular}
\end{center}
\caption{Number of messages containing the keyword ``flu'' and a
  number of keywords that might lead to spurious correlations. The
  final row is the number of flu messages without those spurious
  words. \label{tab:flu}}
\end{table}

Figure \ref{fig:kw2} displays regression results for a number of
potential spurious keywords, such as {\sl swine or H1N1}, {\sl shot},
{\sl vaccine}, {\sl season}, and {\sl http} (to heuristically filter
messages that are simply linking to stories about the flu). Because of
the large amount of noise introduced by discussion of the H1N1 virus,
we have filtered those terms from all results in this figure.  Table
\ref{tab:flu} shows the total number of messages matching each of the
queries.

We make two observations concerning Figure \ref{fig:kw2}. First,
notice that removing messages containing the terms ``swine'' and
``H1N1'' greatly improves correlation on both the training and
evaluation data over using the query ``flu'' alone (training
correlation improves from .93 to .97, evaluation improves from .84 to
.91). 

Second, notice that removing the other spurious terms does not
obviously result in a better fit of the data. In fact, the training
correlation declines by .02, and the evaluation correlation improves
by .01. This result emphasizes the need to {\sl explicitly} test for
robustness in the presence of false alarms, since other measures do
not penalize these spurious terms. We propose such a measure in
Section \ref{sec:simulation}.

\section{Filtering spurious matches by supervised learning}
\label{sec:classifier}
We propose a first step to mitigate the spurious message problem by
training a document classifier to label whether a message is reporting an
ILI-related event or not. This is related to problems such as {\sl
  sentiment analysis} \citep{pang08opinion} and {\sl textual
  entailment} \citep{giampiccolo07third}, which in their most general
form can be quite difficult due to the ambiguities and subtleties of
language. We limit this difficulty somewhat here by only considering
documents that have already matched the hand-chosen ILI-related terms
{\sl flu, cough, headache, sore throat}. The classifier then
calculates a probability that each of these messages is in fact
reporting an ILI symptom.

We train a bag-of-words document classifier using logistic regression
to predict whether a Twitter message is reporting an ILI symptom. Let
$y_i$ be a binary random variable that is 1 if document $d_i$ is a
positive example and is 0 otherwise. Let $\x_i = \{x_{ij}\}$ be a
vector of observed random values, where $x_{ij}$ is the number of
times word $j$ appears in document $i$. We estimate a logistic
regression model with parameters $\theta$ as:
\begin{equation}
p(y_i=1|\x_i; \theta) = \frac{1}{1 + e^{(-\x_i \cdot \theta)}}
\end{equation}
We learn $\theta$ using L-BFGS gradient descent \citep{liu89limited} as
implemented in the MALLET machine learning
toolkit\footnote{\url{http://mallet.cs.umass.edu}}.

\subsection{Combining filtering with regression}
We consider two methods to incorporate the classifier into the
regression model in Equation \ref{eq:regression}.  The first method,
which we term {\bf soft classification}, computes the {\sl expected
  fraction} of positively classified documents as
\begin{equation}
\label{eq:soft}
Q_s(W,D) = \frac{\sum_{d_i \in D_W}p(y_i = 1 | \x_i; \theta)}{|D|}
\end{equation}

This procedure can be understood as weighting each matched document in
$D_W$ by the probability that it is a positive example according to
the classifier. 

The second method, which we term {\bf hard classification}, simply uses
the predicted label for each document, ignoring the class
probability. For the binary case, this simply counts the number of
documents for which the probability of the positive class is greater
than 0.5:

\begin{equation}
\label{eq:hard}
Q_h(W,D) = \frac{\sum_{d_i \in D_W}\mathbf{1}(p(y_i = 1 | \x_i; \theta) > 0.5)}{|D|}
\end{equation}

For both methods, we substitute $Q(W,D)$ in Equation
\ref{eq:regression} with the corresponding classification quantity
from Equation \ref{eq:soft} or \ref{eq:hard}.

\begin{table*}[t]
\begin{center}
\begin{tabular}{|c|}
\hline {\bf Positive Examples} \\ \hline Headache, cold, sniffles, sore
throat, sick in the tummy.. Oh joy !! :' ( \\ me too... i have a
headache my nose is stopped up and it hurts to swallow :/\\ im dying ,
got flu like symptoms, had to phone in ill today coz i was in yest n
was ill and was making mistakes :(\\ \hline {\bf Negative
  Examples}\\ \hline swine flu actually has nothing to do with
swine. \#OMGFACT to the point where they tried to rename the
virus\\ Links between food, migraines still a mystery for headache
researchers \url{http://ping.fm/UJ85w}\\ are you eating fruit
breezers. those other the yummy ones haha. the other ones taste like
well, cough drops haha.\\ \hline
\end{tabular}
\caption{Six Twitter messages labeled as positive or negative examples
  of an ILI-related report. A total of 206 messages were labeled to
  train the classifier of Section
  \ref{sec:classifier}. \label{tab:classification_examples}}
\end{center}
\end{table*}
 
\begin{table}[t]
\begin{center}
\begin{tabular}{|c|c|c|c|}
\hline
{\bf Accuracy} & {\bf F1} & {\bf Precision} & {\bf Recall}\\
\hline
84.29 (1.9) & 90.2 (1.5) & 92.8 (1.8) & 88.1 (2.0)\\
\hline
\end{tabular}
\caption{Results of 10-fold cross validation on the message
  classification task, with standard errors in
  parentheses. \label{tab:classification}}
\end{center}
\end{table}

\begin{figure}[t]
\centering \includegraphics[width=6in]{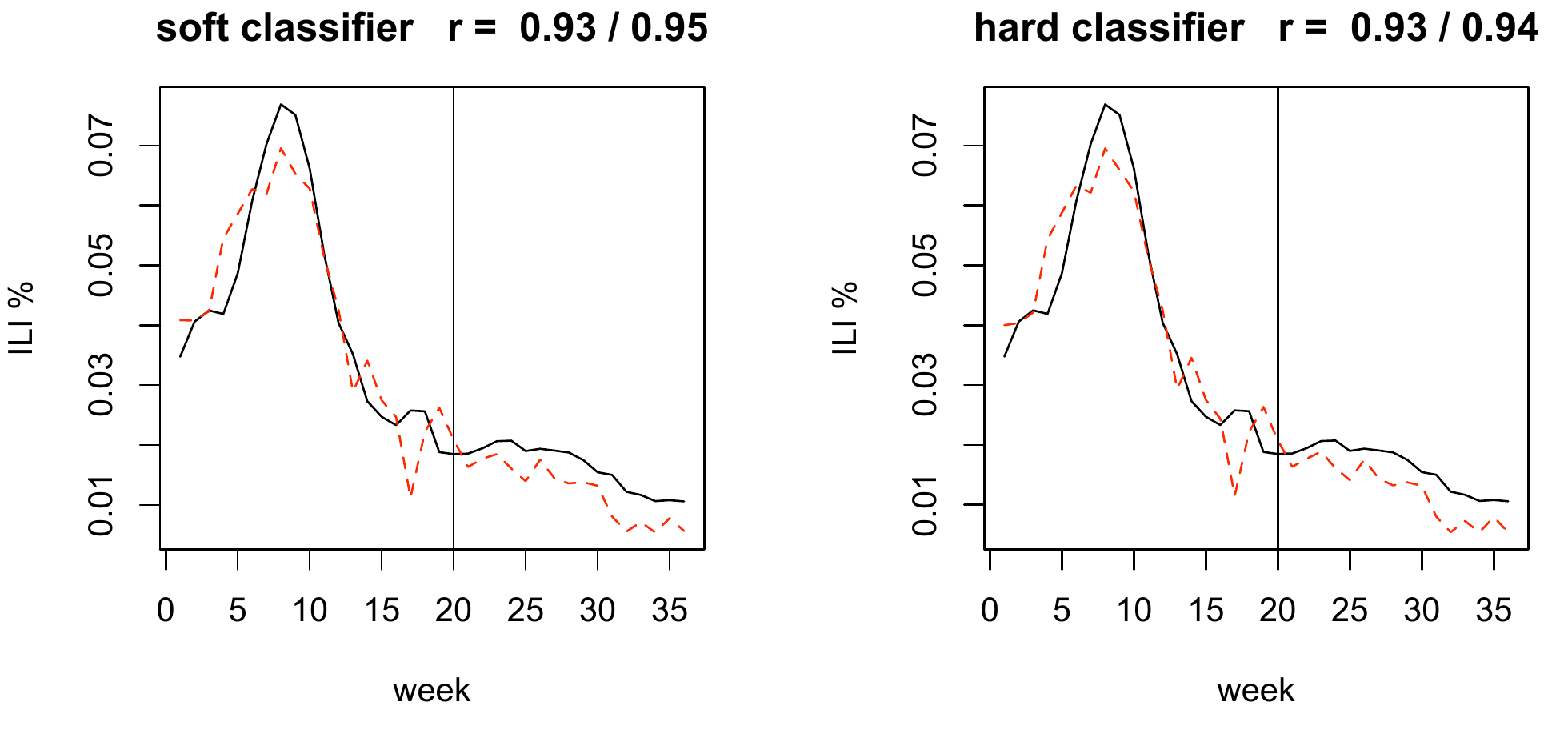}
\caption{Correlation results using both classification strategies to
  filter spurious messages from the query {\sl (flu or cough or
    headache or sore throat}). Filtering does not significantly affect
  the correlation results obtained by the original query, as seen by
  comparing with the final graph in Figure
  \ref{fig:kw}.  \label{fig:classifier}}
\end{figure}

\subsection{Filtering Results}
To generate labeled data for the classification model of Section
\ref{sec:classifier}, we searched for Twitter messages containing any
of {\sl flu, cough, headache, sore throat}, making sure the messages
were posted outside of the date range of the previously collected
messages. This resulted in 206 messages, which were manually
categorized into 160 positive examples and 46 negative
examples. (Examples are shown in Table
\ref{tab:classification_examples}.) Results of 10-fold
cross-validation on this data are shown in Table \ref{tab:classification}.

Figure \ref{fig:classifier} shows the results of both classification
strategies. Note that the initial set of documents are collected by
using the query of the final graph in Figure \ref{fig:kw} ({\sl flu or
  cough or headache or sore throat}). By comparing these two graphs
with the corresponding one in Figure \ref{fig:kw}, we can see that all
three methods perform similarly on this data. This suggests that
filtering out spurious messages does not hurt performance, and that
there likely exists a true underlying correlation between Twitter
messages and ILI rates. The fact that filtering does not improve
performance may suggest that there are no significant spikes in
messages that would lead to a false alarm in the evaluation data.

\section{Evaluating false alarms by simulation}
\label{sec:simulation}
In this section, we evaluate how well the methods proposed in the
previous section filter spikes in spurious messages. Because false
alarms are by definition rare events, it is difficult to use existing
data to measure this. Instead, we propose simulating a false alarm as
follows:

\begin{itemize}
\item We first sample 1,000 messages deemed to be
  spurious.\footnote{We do this by searching for messages containing
    {\sl flu or cough or headache or sore throat} sent by users that were
    news services (e.g., Reuters News). Additionally, we searched for
    messages containing {\sl associated press, AP}, or {\sl health
      officials}.}
\item Next, we sample with replacement an increasing number of these
  spurious messages and add them to the original message set. The
  resulting dataset will add different numbers of spurious messages to
  different weeks.
\item Finally, we use the same trained regression models from Figure
  \ref{fig:classifier} to estimate the ILI rates for each of these
  synthetic datasets.
\end{itemize}

\begin{figure}[t]
\centering \includegraphics[width=5in]{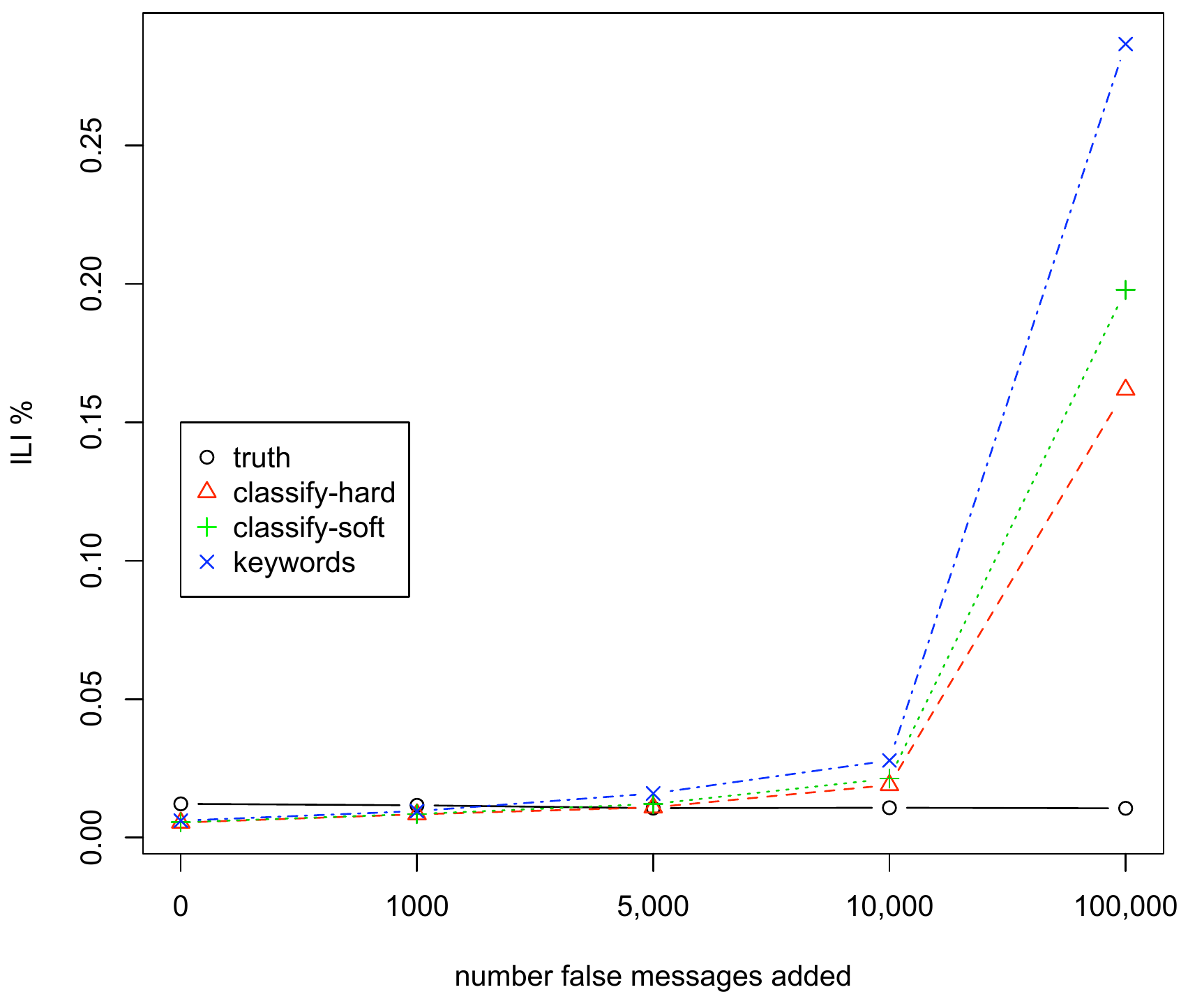}
\caption{Results of the false alarm simulation
  experiment. Corresponding mean-squared errors are {\bf
    keywords=.077, classify-soft=.035, classify-hard=.023}. Hard
  classification appears to be more robust than soft classification in
  the presence of spurious messages, although all approaches are
  overwhelmed when the number of spurious messages reaches
  100,000.  \label{fig:alarm}}
\end{figure}

Figure \ref{fig:alarm} shows the result of this approach for both
classification methods as well as the original keyword based method
(which does no filtering). We simulate weeks 32-36, which have been
augmented with 0, 1000, 5000, 10,000, and 100,000 spurious messages.
We make two observations of these results. First, hard classification
appears to do a better job than soft classification, most likely
because removing any document with probability below 0.5 results in a
much more aggressive filter. Second, it is clear that none of the
methods are adequate under extreme conditions. An additional 100,000
spurious messages overwhelms all approaches and produces an invalid
spike in ILI estimates. We leave investigation of further improvements
for future work.

\section{Related Work}

\citenoun{lampos10tracking} perform a similar analysis of Twitter
message to track influenza rates in the U.K. They learn a
``flu-score'' for each document by learning weights for each word by
their predictive power on held-out data. Using a set of 41 hand-chosen
``markers'', or keywords, they obtain a correlation of 92\% with
statistics reported by the U.K.'s Health Protection
Agency. Additionally, they obtain a correlation of 97\% by using
automated methods to select additional keywords to track (73 in
total), similar to the methodology of
\citenoun{ginsberg09detecting}. In this paper, we have presented a
simpler scheme to track flu rates and have found a comparable level of
correlation as in \citenoun{lampos10tracking}. A principal distinction
of this paper is our attempt to address the issue of false alarms
using supervised learning.

In an earlier version of our work \citep{culotta10towards}, we perform
a similar analysis as in \citenoun{lampos10tracking}, also
experimenting with automated methods to select keywords to track. We
also report an improved correlation by using a classifier to filter
spurious messages. In this paper, we have used similar techniques on a
much larger dataset (570 million vs. 500K). We have also more closely
evaluated the impact of false alarms on these types of methods.

This paper, as well as \citenoun{lampos10tracking} and
\citenoun{culotta10towards}, are similar in methodology to
\citenoun{ginsberg09detecting}, who track flu rates over five years by
mining search engine logs, obtaining a 97\% correlation with ILI rates
on evaluation data. Extending these methods to Twitter, blogs, and
other publicly available resources has the benefit of measuring how
different data sources affect the quality of predictions.

 \citenoun{corley10text} track flu rates by examining the proportion
 of blogs containing two keywords ({\sl influenza} and {\sl flu}),
 obtaining a correlation of 76\% with true ILI rates. It is possible
 that the brevity of Twitter messages make them more amenable to
 simple keyword tracking, and that more complex methods are required
 for blog data.

\section{Conclusions and Future Work}
In this paper, we have provided evidence that relatively simple
approaches can be used to track influenza rates from a large number of
Twitter messages, with the best method achieving 95\% correlation on
held-out data. We have also proposed a supervised learning approach to
reduce the burden of false alarms, and through simulation experiments
we have measured the robustness of this approach. These results
suggest that while document classification can greatly limit the impact of
false alarms, further research is required to deal with extreme cases
involving a large number of spurious messages.

In future work, we plan to consider more sophisticated language
processing approaches to document filtering. Additionally, we plan to
investigate temporal models to further improve forecasting accuracy.

\section{Acknowledgements}
We would like to thank Brendan O'Connor from Carnegie Mellon
University for providing access to the Twitter data and Troy
Kammerdiener of Southeastern Louisiana University for helpful
discussions in early stages of this work. This work was supported in
part by a grant from the Research Competitiveness Subprogram of the
Louisiana Board of Regents, under contract \#LEQSF(2010-13)-RD-A-11.

\bibliographystyle{plainnat} \bibliography{flu_tr}
\end{document}